\documentclass[sigconf,nonacm,10pt]{acmart}

\usepackage{setspace}
\usepackage{hyperref}
\usepackage[most]{tcolorbox}
\usepackage{amsmath}
\usepackage{enumitem}
\usepackage{float}
\usepackage{fontawesome6}
\usepackage{array}
\usepackage{pifont}
 \usepackage{booktabs}
\usepackage{ragged2e}
\usepackage{tabularx}
\usepackage{colortbl}
\usepackage{xcolor}
\usepackage{etoolbox}
\usepackage{comment}
\usepackage{scalefnt}
\usepackage{bm}
\microtypesetup{
    protrusion=true,
    expansion=true
}

\newcommand{\stitle}[1]{\vspace{6pt}\noindent\textbf{#1}}

\newcommand{\connicon}{%
	\raisebox{0ex}{\scalebox{1}{\faLink}}%
}

\newcommand{\keyicon}{%
	\raisebox{0.2ex}{\scalebox{0.8}{\faStar}}%
}

\newcommand{\keyitemicon}{%
	\raisebox{-0.2ex}{\scalebox{1}{\faSquareCaretRight}}%
}

\newcommand{\takeawayicon}{%
	\raisebox{-0.3ex}{\scalebox{1.1}{\faBookmark[regular]}}%
}

\newcommand{\takeawayitemicon}{%
	\raisebox{-0.2ex}{\scalebox{0.9}{\faBookmark}}%
}

\newlist{keyitemize}{itemize}{1}
\setlist[keyitemize]{%
	leftmargin=*,
	itemsep=6pt,
	parsep=0pt,
	label=\keyitemicon
}

\newlist{takeawayitemize}{itemize}{1}
\setlist[takeawayitemize]{%
	leftmargin=*,
	itemsep=1pt,
	parsep=0pt,
	label=\takeawayitemicon
}

 \definecolor{KeyBg}{HTML}{F3F7FC}
 \definecolor{KeyTitle}{HTML}{244A73}

\newtcolorbox{keybox}[1][]{%
    scale=1,
	enhanced,
	tile,
	title= {\textbf{\keyicon \: Key Insights}},
	colback=KeyBg,
	colbacktitle=KeyTitle,
	fonttitle=\large\sffamily\bfseries,
	top=12pt,
	bottom=16pt,
	left=2pt,
	right=9pt,
	toptitle=3pt,
	bottomtitle=3pt,
	lefttitle=3pt,
    before skip=0pt,
    after skip=2pt,
	#1
}

\definecolor{TakeTitle}{HTML}{244A73}
\definecolor{TakeBg}{HTML}{F2F6FA}

 \newtcolorbox{takeawaybox}[2][]{%
    float=htbp,
    tile,
	colback=TakeBg,
	colbacktitle=TakeTitle,
	fonttitle=\large\sffamily,
    title={\textbf{\takeawayicon \: Main Takeaway}\
    \textnormal\sffamily{%
        {\color{white}\normalsize
        [Section~{\hypersetup{linkcolor=white}\ref{#2}}]}}},
	top=4pt,
	bottom=2pt,
	left=4pt,
	right=6pt,
	toptitle=1pt,
	bottomtitle=1pt,
	lefttitle=3pt,
    before skip=4pt,
    after skip=4pt,
	#1
}

\hypersetup{
	colorlinks,%
	citecolor=black,%
	filecolor=black,%
	linkcolor=black,%
	urlcolor=black,%
	pdftitle={Human-Data Interaction, Exploration, and Visualization in the AI Era: Challenges and Opportunities},
	pdfauthor={Jean-Daniel Fekete, Yifan Hu, Dominik Moritz, Arnab Nandi, Senjuti Basu Roy, Eugene Wu, Nikos Bikakis, George Papastefanatos, Panos K. Chrysanthis, Guoliang Li, Lingyun Yu},
	pdfsubject={Human-Centered AI, Visual Analytics, Human-AI Interaction, Multimodal Analytics},
	pdfkeywords={Human-in-the-Loop, Human-Centered AI, Human–AI Collaboration, Generative Visualization, Visual Analytics, AI-Guided Exploration, Interactive ML, Multimodal Visualization, Progressive and Scalable Visual Analytics, Foundation Models, Mixed-Initiative Interaction, AI-Assisted Data Analysis, Narrative Visualization, LLMs, VLMs, Human-Centered Databases}
}

\definecolor{myboxcolor2}{HTML}{9EC6EA}
\colorlet{CHALHeader}{myboxcolor2!35!black}
\colorlet{CHALLeftBg}{red!3!white}
\definecolor{CHALMiddleBg}{HTML}{FFFFFF}
\definecolor{CHALOppBg}{HTML}{FFFFFF}

\colorlet{CHALTagBg}{red!14!white}
 \definecolor{CHALRule}{HTML}{DDE3E6}
\colorlet{CHALChallengeTitle}{red!50!black}
\newcommand{\CHALtag}[1]{{\setlength{\fboxsep}{2.3pt}%
  \colorbox{CHALTagBg}{\textcolor{CHALChallengeTitle}{\sffamily\fontsize{8}{8.5}\selectfont\bfseries #1}}}}
 \newcommand{\CHALsections}[1]{%
    {%
    \hypersetup{linkcolor=CHALHeader}%
    \fontsize{8}{9.3}\selectfont
    \color{CHALHeader}#1%
  }%
}

\author{Jean-Daniel Fekete}
\affiliation{\institution{Univ. Paris-Saclay \& Inria}\country{France}}

\author{Yifan Hu}
\affiliation{\institution{Northeastern Univ.}\country{USA}}

\author{Dominik Moritz}
\affiliation{\institution{Carnegie Mellon Univ. \& Apple}\country{USA}}

\author{Arnab Nandi}
\affiliation{\institution{Ohio State Univ.}\country{USA}}

\author{Senjuti Basu Roy}
\affiliation{\institution{New Jersey Inst. of Technology}\country{USA}}

\author{Eugene Wu}
\affiliation{\institution{Columbia Univ.}\country{USA}}

\author{Nikos Bikakis}
\affiliation{\institution{Hell. Med. Univ. \& Archimedes/Athena RC}\country{Greece}}

\author{George Papastefanatos}
\affiliation{\institution{Athena RC}\country{Greece}}

\author{Panos K. Chrysanthis}
\affiliation{\institution{Univ. of Pittsburgh}\country{USA}}

\author{Guoliang Li}
\affiliation{\institution{Tsinghua Univ.}\country{China}}

\author{Lingyun Yu}
\affiliation{\institution{Xi'an Jiaotong-Liverpool Univ.}\country{China}}

\begin{document}

\pagestyle{empty}

\title{\vspace{-6pt}Human--Data Interaction, Exploration, and Visualization Systems in the AI Era: Challenges and Opportunities}

\font\CHALauthorfont=phvr8t at 12pt
\font\CHALaffilfont=phvr8t at 10pt
\makeatletter
\renewcommand{\@mkauthors}{%
  \global\setbox\mktitle@bx=\vbox{%
    \unvbox\mktitle@bx
    \vskip 10pt
    \centering
    {\CHALauthorfont
    Jean-Daniel Fekete\textsuperscript{1}\hspace{1em}
    Yifan Hu\textsuperscript{2}\hspace{1em}
    Dominik Moritz\textsuperscript{3}\hspace{1em}
    Arnab Nandi\textsuperscript{4}\\[2pt]
    Senjuti Basu Roy\textsuperscript{5}\hspace{0.5em}
    Eugene Wu\textsuperscript{6}\hspace{0.5em}
    Nikos Bikakis\textsuperscript{7}\hspace{0.5em}
    George Papastefanatos\textsuperscript{8}\\[2pt]
    Panos K. Chrysanthis\textsuperscript{9}\hspace{0.5em}
    Guoliang Li\textsuperscript{10}\hspace{0.5em}
    Lingyun Yu\textsuperscript{11}\par}
    \vskip 6pt
    {\CHALaffilfont
    \textsuperscript{1}Univ. Paris-Saclay \& Inria, France\hspace{0.5em}
    \textsuperscript{2}Northeastern Univ., USA\hspace{0.5em}
    \textsuperscript{3}Carnegie Mellon Univ. \& Apple, USA\\[2pt]
    \textsuperscript{4}Ohio State Univ., USA\hspace{0.5em}
    \textsuperscript{5}New Jersey Inst. of Technology, USA\hspace{0.5em}
    \textsuperscript{6}Columbia Univ., USA\\[2pt]
    \textsuperscript{7}Hell. Med. Univ. \& Archimedes/Athena RC, Greece\hspace{0.5em}
    \textsuperscript{8}Athena RC, Greece\\[2pt]
    \textsuperscript{9}Univ. of Pittsburgh, USA\hspace{0.5em}
    \textsuperscript{10}Tsinghua Univ., China\hspace{0.5em}
    \textsuperscript{11}Xi'an Jiaotong-Liverpool Univ., China\par}
    \vskip 10pt
  }%
}
\makeatother

\begin{abstract}
 The rapid advancement of AI is transforming human--centered systems, with profound implications for human--AI interaction, human--data interaction, and visual analytics.
In the AI era, data analysis increasingly involves large-scale, heterogeneous, and multimodal data that is predominantly unstructured, as well as foundation models such as LLMs and VLMs, which introduce additional uncertainty into analytical processes.
These shifts expose persistent challenges for human--data interactive systems, including perceptually misaligned latency, scalability constraints, limitations of existing interaction and exploration paradigms, and growing uncertainty regarding the reliability and interpretability of AI--generated insights.
Responding to these challenges requires moving beyond conventional efficiency and scalability metrics, redefining the roles of humans and machines in analytical workflows, and incorporating cognitive, perceptual, and design principles into every level of the human--data interaction stack.
This paper investigates the challenges introduced by recent advances in AI and examines how these developments are reshaping the ways users engage with data, while outlining limitations and open research directions for building human--centered AI systems for interactive data analysis in the AI era.
\end{abstract}

\keywords{Human-AI Collaboration, Visual Analytics, Multimodal Analytics, Generative Visualization, AI-guided Exploration, Foundation Models, Trustworthy AI}

 \maketitle


\section{Introduction}
\label{sec:intro}

The growing influence of AI is fundamentally changing how humans interact with, explore, and reason about data, as well as how human--centered systems are designed and implemented. Advances in AI--driven techniques have expanded the scope of interactive data analysis beyond structured tables to include \textit{large-scale, heterogeneous, and multimodal data} such as text, images, audio, video, and visually rich documents. As a result, data analysis increasingly involves information that is unstructured, weakly indexed, and costly to inspect, which challenges traditional \textit{query--then--explore workflows} that rely on well-defined schemas, stable metadata, and predictable access patterns. These shifts place new demands on human--centered systems.

In practice, users must \textit{iteratively explore data, reformulate queries, and verify intermediate results under uncertainty}, often without a clear understanding of what the data contains or which questions can be meaningfully asked.
At the same time, \textit{interactive analysis imposes strict perceptual latency constraints}.
Interaction delays can affect exploratory behavior and
visual search~\cite{LiuH14,BattleLatency20},
potentially disrupting analytical reasoning and the
effectiveness of human--AI collaboration.
Taken together, these challenges
reveal fundamental limitations in existing system architectures, data management techniques, interaction paradigms, and visualization approaches.
Addressing these challenges requires rethinking interactive data systems as \textit{end--to--end, human--centered pipelines}. Rather than treating data management, AI models, interfaces, and visualization as loosely coupled components, future systems must be \textit{co-designed} to support iterative, guidance-driven, and mixed-initiative exploration.

\textit{Latency} and \textit{scalability} emerge as cross-cutting constraints. They shape not only system performance, but also user behavior, cognitive load, and the feasibility of exploratory workflows. Supporting interaction at the \textit{speed of human thought} requires new abstractions, query processing methods, indexing schemes, and approximation strategies that account for interface structure, human perception, and interaction context.

The growing reliance on foundation models (e.g., LLMs, VLMs) and other deep learning models (e.g., GANs, GNNs) also introduces new concerns, including non-determinism, error propagation, bias, and the cost of verification. These issues reinforce the necessity of \textit{keeping humans   involved in the analytical loop}, with explicit support for oversight, provenance, and trust calibration.

Interfaces themselves are also evolving. \textit{Next-generation interfaces} increasingly move toward multimodal, action-oriented environments that integrate natural language interaction, gestures, immersive technologies such as augmented and virtual reality (AR/VR), and continuous  feedback. These interfaces enable more direct and intuitive engagement with complex data, but they also impose new requirements on underlying systems, including \textit{real-time responsiveness, progressive result refinement}, and \textit{adaptive guidance}. In such settings, interaction becomes tightly coupled with perception and action, further amplifying the need for co-design across systems, interfaces, and visualization.

In this evolving landscape, \textit{visualization} plays an increasingly central role.
Visualization is  an \textit{active system component} that supports exploration, query reformulation, guidance, verification, and feedback. Progressive, adaptive, and narrative-driven visual representations can help users navigate uncertainty, focus attention, and assess the reliability of AI--assisted results.

\stitle{Summary.}
	The AI era ecosystem requires that human--data interaction systems deliver real-time, cognition-aware performance and scalability, effective multimodal data handling, and advanced visual analytics, while simultaneously supporting user guidance, progressive sensemaking, and trustworthy analysis under AI uncertainty.

Drawing on perspectives from databases, AI, information visualization, human--computer interaction, and computer graphics, this paper presents the key challenges and   research directions for building scalable, reliable, and human--centered interactive data systems in the AI era.
Throughout this paper, we use human--data interaction in a broad sense, encompassing how people access, explore, analyze, understand, and manipulate data, as well as the involved systems and mechanisms.
Each section presents the perspective and insights of an author, as follows:
Eugene Wu~(Sec.~\ref{sec:eugene}), Dominik Moritz~(Sec.~\ref{sec:dominik}),
Jean-Daniel Fekete~(Sec.~\ref{sec:jean_daniel}), Arnab Nandi~(Sec.~\ref{sec:arnab}),
 Senjuti Basu Roy~(Sec.~\ref{sec:senjuti}), and Yifan Hu~\mbox{(Sec.~\ref{sec:yifan})}.
The first author contributed equally to the manuscript content, while the last five authors coordinated the preparation
and enrichment of the manuscript.

\stitle{Terminology and Definitions.}
 Here, \textit{latency} is the time to a specified system response, whereas
\textit{responsiveness} concerns whether that response arrives in time to
support the interaction. \textit{Scalability} concerns how performance,
preparation cost, and resource use change as data, workload, or model scale
grows. \textit{Guidance} is
\linebreak

\begin{keybox}
	\begin{keyitemize}

		\item
		\textbf{The AI era introduces new, tightly coupled requirements for human--data interaction systems.}
		Systems must deliver perceptually aligned responsiveness and scalability, natively handle multimodal and unstructured data, support guidance and emerging interaction and exploration paradigms, leverage visualization capabilities, and remain robust to AI uncertainty via progressive feedback and trustworthy analysis.

		\item
		\textbf{Latency and scalability are fundamental barriers.}  Challenges such as multimodal data, next-generation interfaces, embeddings, complex interactions, and high-velocity streams  require rethinking core data management techniques to meet perceptual latency constraints and support the development of scalable and reliable human--centered systems. DL models must also scale to large training datasets.

		\item
		\textbf{End--to--end system design is essential.}
		Interfaces must evolve toward multimodal, action-oriented, and adaptive environments that integrate modalities  like NL, gestures, narrative-rich generative visualizations, and AR/VR.
		Supporting these capabilities, as well as low-latency performance and scalability, requires co-design across system architectures, data operations, AI components, interfaces, and visualization.

		\item
		\textbf{AI models are powerful but require human--in--the--loop oversight.}
		AI--driven techniques can enable guidance-driven exploration, semantic understanding, data-oriented and exploratory tasks, and improved system usability. However, their brittleness, error-proneness, unpredictability, and biases underscore that human involvement is essential.


        \item
        \textbf{Generative and adaptive methods can augment interactive visualization.}
        Progressive, narrative, and learned visual layouts guide attention, enhance interpretability, and mediate trust in AI--driven analytics.

		\item
		\textbf{Sustained cross-disciplinary collaboration is crucial.}
		The diversity of emerging challenges underscores the need for close collaboration among the systems, databases, AI, information visualization, HCI, computer graphics, and cognitive science communities.

	\end{keyitemize}
\end{keybox}

\begin{table*}[t]
	{
		\centering
		\caption{Challenges and Opportunities Landscape.}
		\label{tab:shared-challenges}
		\begingroup
		\setlength{\tabcolsep}{3pt}
		\fontsize{10}{11.5}\selectfont

		\arrayrulecolor{CHALRule}

		\makebox[\textwidth][c]{\scalebox{1}{%
				\begin{tabularx}{\textwidth}{@{}
						>{\columncolor{CHALLeftBg}[3pt][3pt]
							\RaggedRight\hyphenpenalty=10000\arraybackslash}p{0.238\textwidth}
						>{\columncolor{CHALMiddleBg}[3pt][3pt]
							\RaggedRight\arraybackslash}X
						>{\columncolor{CHALOppBg}[3pt][3pt]
							\RaggedRight\arraybackslash}X
						@{}}

					\rowcolor{CHALHeader}[3pt][3pt]
					\rule[-5pt]{0pt}{17pt}\textcolor{white}{\sffamily\bfseries\fontsize{9.5}{11}\selectfont Challenge}
					& \textcolor{white}{\sffamily\bfseries\fontsize{9.5}{11}\selectfont Aspects}
					& \textcolor{white}{\sffamily\bfseries\fontsize{9.5}{11}\selectfont Opportunities} \\

					\specialrule{0pt}{1pt}{3pt}

					\CHALtag{C1}\enspace
					\textbf{\textcolor{CHALChallengeTitle}{\sffamily Responsiveness}}\par\vspace{2pt}
					Keep interactive feedback aligned with the timescales of human perception and reasoning.
					\textbf{\CHALsections{[Sec.~\ref{sec:eugene}, \ref{sec:dominik}, \ref{sec:jean_daniel}, \ref{sec:arnab}]}}
					&
					coordinated view updates and interface-aware latency \CHALsections{(Sec.~\ref{sec:eugene})};
					delays that disrupt exploration \CHALsections{(Sec.~\ref{sec:dominik})};
					cold-start preparation \CHALsections{(Sec.~\ref{sec:jean_daniel})};
					real-time multimodal/AR feedback \CHALsections{(Sec.~\ref{sec:arnab})}.
					&
					interface-aware optimization \CHALsections{(Sec.~\ref{sec:eugene})};
					approximation, indexing, and prefetching \CHALsections{(Sec.~\ref{sec:dominik})};
					progressive computation/AQP \CHALsections{(Sec.~\ref{sec:jean_daniel})};
					latency-aware multimodal feedback \CHALsections{(Sec.~\ref{sec:arnab})}. \\

					\specialrule{0.4pt}{2pt}{2pt}

					\CHALtag{C2}\enspace
					\textbf{\textcolor{CHALChallengeTitle}{\sffamily Scalability}}\par\vspace{2pt}
					Sustain interactive analysis as data, workloads, and models grow in scale.
					\textbf{\CHALsections{[Sec.~\ref{sec:eugene}--\ref{sec:yifan}]}}
					&
					complex relational workloads \CHALsections{(Sec.~\ref{sec:eugene})};
					large model datasets \CHALsections{(Sec.~\ref{sec:dominik})};
					dense/high-cardinality data \CHALsections{(Sec.~\ref{sec:jean_daniel})};
					large multimodal collections \CHALsections{(Sec.~\ref{sec:arnab})};
					repeated extraction/model calls \CHALsections{(Sec.~\ref{sec:senjuti})};
					large learned-layout workloads \CHALsections{(Sec.~\ref{sec:yifan})}.
					&
					scalable execution and verified synthesis \CHALsections{(Sec.~\ref{sec:eugene})};
					reusable optimization abstractions \CHALsections{(Sec.~\ref{sec:dominik})};
					progressive/AQP exploration \CHALsections{(Sec.~\ref{sec:jean_daniel})};
					approximate multimodal querying \CHALsections{(Sec.~\ref{sec:arnab})};
					reusable extracted attributes \CHALsections{(Sec.~\ref{sec:senjuti})};
					scalable learning-based visualization \CHALsections{(Sec.~\ref{sec:yifan})}. \\

					\specialrule{0.4pt}{2pt}{2pt}

					\CHALtag{C3}\enspace
					\textbf{\textcolor{CHALChallengeTitle}{\sffamily Guidance \& User Control}}\par\vspace{2pt}
					Help users discover, formulate, and refine analytical questions while preserving user control.
					\textbf{\CHALsections{[Sec.~\ref{sec:eugene}, \ref{sec:dominik}, \ref{sec:arnab}, \ref{sec:senjuti}, \ref{sec:yifan}]}}
					&
					agent-generated interfaces and user expectations \CHALsections{(Sec.~\ref{sec:eugene})};
					emergent comparisons \CHALsections{(Sec.~\ref{sec:dominik})};
					multimodal query-formulation barriers \CHALsections{(Sec.~\ref{sec:arnab})};
					query reformulation \CHALsections{(Sec.~\ref{sec:senjuti})};
					user-constrained layouts \CHALsections{(Sec.~\ref{sec:yifan})}.
					&
					inspectable, personalized interactions \CHALsections{(Sec.~\ref{sec:eugene})};
					mixed-initiative subset exploration \CHALsections{(Sec.~\ref{sec:dominik})};
					multimodal autocompletion and evidence-linked cues \CHALsections{(Sec.~\ref{sec:arnab})};
					attribute-driven reformulation and mental models \CHALsections{(Sec.~\ref{sec:senjuti})};
					constraint-aware layouts and attention guidance \CHALsections{(Sec.~\ref{sec:yifan})}. \\

					\specialrule{0.4pt}{2pt}{2pt}

					\CHALtag{C4}\enspace
					\textbf{\textcolor{CHALChallengeTitle}{\sffamily Reliability \& Appropriate Reliance}}\par\vspace{2pt}
					Enable users to assess system and AI outputs and rely on them appropriately under uncertainty and verification cost. \linebreak
					\textbf{\CHALsections{[Sec.~\ref{sec:eugene}, \ref{sec:arnab}, \ref{sec:senjuti}, \ref{sec:yifan}]}}
					&
					hidden data flows and shared-state conflicts \CHALsections{(Sec.~\ref{sec:eugene})};
					uncertain multimodal suggestions and costly verification \CHALsections{(Sec.~\ref{sec:arnab})};
					uncertain or biased LLM attributes \CHALsections{(Sec.~\ref{sec:senjuti})};
					limits of preference scores and generated narratives \CHALsections{(Sec.~\ref{sec:yifan})}.
					&
					provenance, policy enforcement, and safe branching \CHALsections{(Sec.~\ref{sec:eugene})};
					source-linked verification and correction \CHALsections{(Sec.~\ref{sec:arnab})};
					provenance, uncertainty, and trust calibration \CHALsections{(Sec.~\ref{sec:senjuti})};
					separate evaluation of preference, task performance, and faithfulness \CHALsections{(Sec.~\ref{sec:yifan})}. \\

					\specialrule{0.4pt}{2pt}{2pt}

					\CHALtag{C5}\enspace
					\textbf{\textcolor{CHALChallengeTitle}{\sffamily Effective Visual Representation}}\par\vspace{2pt}
					Keep visual representations interpretable and analytically useful as data and outputs grow more complex.
					\textbf{\CHALsections{[Sec.~\ref{sec:eugene}, \ref{sec:dominik}, \ref{sec:jean_daniel}, \ref{sec:arnab}, \ref{sec:yifan}]}}
					&
					limited relational visualization abstractions \CHALsections{(Sec.~\ref{sec:eugene})};
					fixed dashboards and large-data comparisons \CHALsections{(Sec.~\ref{sec:dominik})};
					skew, dense regions, and many categories \CHALsections{(Sec.~\ref{sec:jean_daniel})};
					verbose VLM guidance \CHALsections{(Sec.~\ref{sec:arnab})};
					learned layouts and narratives \CHALsections{(Sec.~\ref{sec:yifan})}.
					&
					richer relational interfaces \CHALsections{(Sec.~\ref{sec:eugene})};
					responsive editable views \CHALsections{(Sec.~\ref{sec:dominik})};
					adaptive/hybrid visualization \CHALsections{(Sec.~\ref{sec:jean_daniel})};
					compact evidence-linked cues \CHALsections{(Sec.~\ref{sec:arnab})};
					constrained learned layouts and data-grounded narratives \CHALsections{(Sec.~\ref{sec:yifan})}. \\

					\specialrule{0.9pt}{3.5pt}{0pt}

				\end{tabularx}
		}}
		\endgroup
	}
\end{table*}

\noindent
system support for deciding what to inspect, ask,
or do next; \textit{user control} is the ability to steer, revise, or reject
such assistance. \textit{Progressive feedback} exposes intermediate results
that are refined during interaction.

AI \textit{uncertainty} concerns instability or limited confidence in
AI-assisted results due to non-determinism or model limitations.
\textit{Reliability} concerns the dependability of system or AI outputs,
whereas \textit{trust} is the user's attitude toward that dependability and
\textit{reliance} concerns acting on those outputs.
\textit{Trust calibration} describes how well trust matches
system capabilities and limitations; \textit{appropriate reliance}
occurs when users' reliance on system outputs is consistent with
those capabilities and the task~\cite{LeeSee2004}.
\textit{Verification cost} is the time and effort required to assess an output.
Verification cost (Sec.~\ref{sec:arnab}), uncertainty in AI-assisted
exploration (Sec.~\ref{sec:senjuti}), and explanation and accountability
(Sec.~\ref{sec:disc}) can all affect appropriate reliance. We use
\textit{challenge} for a recurring limitation or requirement and
\textit{opportunity} for a research direction that may address it.

AI-specific \textit{failure modes} can be grouped into three categories:
\textit{grounding failures}, such as hallucinated or unsupported outputs;
\textit{output instability}, including non-determinism
and prompt sensitivity; and \textit{systematic distortions}, such as bias and
uneven coverage across data groups.
These failure modes may overlap and affect different
stages of the human--data interaction pipeline,
requiring distinct verification and mitigation strategies.

\enlargethispage{\baselineskip}
\stitle{Unifying Challenge Framework.}
Across the paper, we identify five recurring challenges:
\textbf{C1}: \textit{Responsiveness};
\textbf{C2}: \textit{Scalability};
\textbf{C3}: \textit{Guidance \& User Control};
\textbf{C4}: \textit{Reliability \& Appropriate Reliance}; and
\textbf{C5}: \textit{Effective Visual Representation}.
Table~\ref{tab:shared-challenges} condenses how these challenges appear in
Sections~\ref{sec:eugene}--\ref{sec:yifan} and the corresponding research
opportunities.
Section~\ref{sec:disc} expands on the challenges and opportunities outlined in the second and third columns, and discusses how they interact, compound, and involve trade-offs.

 \stitle{Further Reading.}
For further details, readers can refer to representative
literature organized around surveys and reviews, research perspectives, and benchmarks.

\begin{itemize}[leftmargin=*, itemsep=4pt, topsep=0pt, parsep=0pt]

\item
\textit{Human-Centered Data Management.}
Reviews of human--data interaction systems are provided
in~\cite{WuC25,Battle2020}, while data-management support
for visualization is surveyed in~\cite{QinLTL20}.
Research perspectives discuss emerging challenges and opportunities
at the intersection of data management and visualization
in~\cite{blog,bigvissurvey,AndrienkoADFFIK20}.
Work on scalability and progressive visualization is covered
in~\cite{richer2024scalability,UlmerAFKM24},
while design--performance integration and large-scale visualization
are discussed in~\cite{Wu2017CombiningDA,Bikakis2022}.
Finally, benchmarking studies examine workloads and evaluation
approaches for interactive data exploration and visual
analysis~\cite{jiang2018evaluating,BattleEACSZBFM20,EichmannZBK20}.

\item
\textit{Visual Analytics.}
Visual insight mining techniques are surveyed
in~\cite{LIAN2025100271}, while broader research challenges
and directions in visual analytics are discussed
in~\cite{WuDCLKMMSVW23}.
Complementary work examines visualization design processes
in~\cite{WalnyFWKKCW20} and human-centered methodological
aspects in~\cite{ADAMAKIS2025100269}.

\item
\textit{AI and Visual Analytics.}
Surveys on AI and ML techniques for visualization are provided
in~\cite{WuWSMCZZQ22,WangCWQ22}.
Visual analytics for understanding and interpreting ML models
is reviewed in~\cite{WangLZ24,YuanCYLXL21,HohmanKPC19}.
Recent work examines generative AI and foundation models
for visualization~\cite{BasoleM24,YeHHWXLZ24},
including emerging approaches and challenges~\cite{YangLWL24}.
The LLM-enabled interaction with visualizations is reviewed in~\cite{Brossier2026StateOf}, while the role of interactive
visualization in human-centered AI tools is discussed
in~\cite{hoque2024}.
Broader human-centered research directions are discussed
in~\cite{AndrienkoAAWR22,ElmqvistK25},
while evaluation benchmarks are presented in~\cite{ChenZXRY25}.

\item
\textit{Human--AI Collaboration.}
Reviews on human-centered AI and human--AI interaction
are provided in~\cite{Shneiderman22} and~\cite{RaeesMLKP24},
while the growing integration of LLMs into HCI research
is reviewed in~\cite{PangLLMCHI25}.
Guidelines for human--AI interaction design are discussed
in~\cite{AmershiWVFNCSIB19}, while collaborative data storytelling
and visualization support are covered in~\cite{LiWQ24}.

\item
\textit{Sensemaking and Cognitive Task Analysis.}
Foundational work examines the cost structure of
sensemaking~\cite{v4_russell1993},
while subsequent research investigates sensemaking
processes and opportunities for supporting analytical
workflows~\cite{v4_pirolli2005}.

\item
\textit{Visual Perception, Literacy, and Decision Making.}
Perception-based visualization studies are surveyed
across tasks~\cite{QuadriPerception22},
while visualization literacy and its assessment are
reviewed in~\cite{VaronaLiteracy26}.
The role of decision-making tasks in visualization
research is examined in~\cite{DimaraDecision22}.

\item
\textit{Interaction Latency and Human Performance.}
Studies examine the effects of latency on visual search
and task complexity~\cite{BattleLatency20},
pointing and steering tasks~\cite{FristonLatency16},
and human--LLM interaction~\cite{TanLatency26}.

\item
\textit{Guidance and User-Adaptive Visual Analytics.}
Guidance in visual analytics is characterized
in~\cite{CenedaGuidance17},
while principles of mixed-initiative interaction
are discussed in~\cite{v4_horvitz1999}.
More recent surveys cover mixed-initiative visual
analytics~\cite{Monadjemi2026MIVA}
and user-adaptive visualization techniques~\cite{YanezAdaptive25}.

\item
\textit{Human--AI Roles, Agency, and Mental Models.}
Recent reviews examine methods for studying users'
mental models of AI~\cite{SanchezMentalModels26},
the roles assigned to LLMs in decision-making~\cite{ChappidiRoles26},
and patterns of human--AI agency in collaborative
tasks~\cite{ZhangAgency25}.
Cognitive challenges in prompt-based interaction
are examined in~\cite{v4_subramonyam2024}.

\item
\textit{Human-Centered Explainability and Evaluation.}
Surveys examine human-centered XAI design
principles~\cite{HughesHCXAI26}
and interfaces for presenting AI
explanations~\cite{CappuccioXAIUI26}.
Affective aspects of explainable AI are discussed
in~\cite{v4_bernardo2023},
while frameworks for designing and evaluating AI
explanations are reviewed in~\cite{CortinasXAI25}.

\item
\textit{Uncertainty and Trust Calibration.}
Uncertainty-aware guidance in visual analytics is surveyed
in~\cite{Musleh2026GuidingThrough},
while approaches to representing AI uncertainty for
human understanding are examined in~\cite{v4_cassenti2023}.
Foundational principles of trust and appropriate reliance
are discussed in~\cite{LeeSee2004},
with recent perspectives on trust calibration in
visualization provided in~\cite{JusufiTrust26}.
\end{itemize}


\section{Interface-System Co-Design}
\label{sec:eugene}
Data management as a discipline is fundamentally about making it easier to see, analyze, and interact with data by designing data models, query abstractions, and execution engines---always with an application class in mind.
Data visualization pursues a closely aligned goal, but centers its focus on on user experience, perceptually-aware design, and interaction techniques.
Systems for human--data interaction sit  at the intersection of both disciplines and must account for both usability, resource, and system constraints, which often conflict with each other.
 However, doing so is difficult at arm's length.
Progress requires not only modeling interfaces and user interactions in ways that make system design tractable, but also leveraging system and query capabilities to expand the interfaces that are possible.

This section summarizes the evolution of how data systems have used increasingly richer models of the interface to inform the system designs.   We further argue that system design need not be reactive to {\it known} application needs, but can be inspiration for  new interface and interaction designs. These arguments are discussed in greater depth in  \cite{WuC25}.

\stitle{Interfaces as SQL-Workloads.} The most direct approach–what we term the \mbox{SQL-workload} approach–is to treat visualization as a SQL-workload generator. This is sensible because most visualization systems indeed translate user interactions into SQL queries, execute those queries in a DBMS, and then update the interface with the query results. For instance, IDEBench~\cite{EichmannZBK20} generates SQL queries inspired by cross-filtering visualizations.  Under this model, improving query execution throughput or latency directly improves the responsiveness of the interfaces, and variations of classic database techniques such as   approximation~\cite{RamanH01,LiWYZ19,DingHCC016,Kraska18},
query prefetching~\cite{doshi2003prefetching,Shneiderman08,BattleCS16}
shared scans~\cite{khan2019flux},
precomputation \cite{PahinsSSC17,AgarwalMPMMS13}, and indexing~\cite{TaoLWBDCS19} have shown success.

\stitle{Interfaces as System Constraints.} \textit{However, this SQL-workload approach does not focus on the interface’s unique set of constraints}. These constraints motivate new and challenging classes of data management problems.
First, as pointed out by Jiang et al.~\cite{jiang2018evaluating}, \textit{simple workload statistics like P95 latency or query throughput do not translate into fluid user experiences}.
A single click may trigger a dozen parallel queries and the user expectation can vary from all queries
responding simultaneously, only the queries ``above the fold'', or only the query for the ``central chart''.
In addition, \textit{when} components of the interface update in response to interactions
matters~\cite{Wu2017CombiningDA,Wu2019DIELIV,Wu2018FacilitatingEW,Fekete2023TransactionalPA}.

 As Chen et al.~\cite{chen2025physical} argue, there is a \textit{Design Dependence} between how interfaces and interactions are specified and the system designs required to support them. In practice, this dependence cuts both ways: systems are often aggressively optimized for a narrow set of visualization and interaction patterns, which in turn biases designers toward interfaces that are easy to build rather than those that are desirable or expressive. As researchers develop increasingly specialized optimizations---each encodes implicit assumptions about the user needs, interface designs, and interaction semantics---the complexity of selecting, configuring, and composing these optimizations grows combinatorially. Much like physical database design, this burden should not be placed on human designers. This can result in a form of ``streetlight effect,'' where interface designs converge around what existing systems easily support, rather than exploring what is possible or effective. This motivates the need for interface-aware system optimizers that can reason jointly about interface structure and system design~\cite{chen2025physical}.

Second, SQL-workloads are only representative of visualizations created \textit{today},
which themselves are limited by what \textit{today’s} systems can support.
For instance, the  majority of visualization systems, and thus the SQL-workloads,
assume that the input is a single table or view.
Yet hiding the underlying join dependencies introduces subtle
semantic errors~\cite{Huang2023AggregationCE,hyde2024measures}.
Techniques based on factorized query execution and semiring
aggregation have been successfully adapted to scale interactive
visual analysis over large join graphs~\cite{Huang2023LightweightMF},
and introduce a variation of view maintenance under incremental changes
to the query, rather than the input data.

 \stitle{Interfaces as First-class Abstractions.}
As a full-stack community, we have the option of introducing new system designs,
and changing the system abstraction to access new performance
frontiers and empowering interfaces with new capabilities.
This can vary from introducing a new API so designers can specify new models of user perception
to inform approximation algorithms~\cite{Alabi016},
replacing traditional request-response prefetching with a communication model inspired
by video streaming~\cite{MohammedWNW20}, and identifying the equivalence between interaction and
provenance~\cite{Psallidas2018ProvenanceFI} to simplify interface development~\cite{Psallidas2018DemonstrationOS}
and advance provenance systems research~\cite{Psallidas2018SmokeFL,Mohammed2023SmokedDuckDS} at the same time.

Data management research naturally translates into new visualization features,
but requires a careful co-design with interface and interactivity requirements.
For instance, the data management community has explored algorithmic techniques to explain analysis results~\cite{gathani2020debugging,roy2015explaining,huang2022reptile,wu2020complaint,wu2013scorpion,abuzaid2018diff},
but careful design, interactive latencies, and simple abstractions are needed for
wide-spread adoption in \textit{any} visualization~\cite{mohammed2025fade}.
As another example, data cube and comparison theory from two decades ago~\cite{i3} can  motivate
new interaction designs to support visual comparison of any combination of charts, marks,
and/or values in a data interface~\cite{wu2022view,vcahier}.

\stitle{Rethinking Visualization Interfaces Altogether.}
It is also valuable to rethink how visualizations are designed, optimized, and implemented.
Today, this requires a unicorn designer who is an expert in the domain,
visualization design and implementation, and system optimization,
or to coordinate hand-offs of a team of designers and developers—known to be fraught with
challenges~\cite{WalnyFWKKCW20}.
What if we could synthesize the interface design (or thousands of designs)
simply from samples of the desired analysis queries~\cite{Chen2021PI2EI,Zhang2017MiningPI}, annotate the
interactions with their latency requirements, and automatically instantiate an end--to--end client-server
architecture that, when pointed to a cloud DBMS, guarantees the desired interactivity~\cite{chen2025physical}?

Data visualization is a young and evolving field, and data management research can also inform new visualization abstractions and theories.
For instance, what is an intermediate representation of the interface and analysis that is easy to write, amenable to automatic optimization, and can be used to rapidly synthesize new interface designs?
Recent work by Chen et al.~\cite{Chen2023DIGTD} proposes a Data Interface Grammar, but this remains an open challenge.
Another open question is the line between data processing and visualization—how much of the visualization and user analysis process can be modeled as ``just a query?''~\cite{wu2024design}.

Finally, as hinted at above, visualization theory that modern visualization
grammars and libraries are predicated on, and fundamentally assumes, that the input is a single table.
Yet relational theory makes it clear that a single-table does not adequately model all data.
Is there a way to extend ``single-table'' visualization theory to a theory of
\textit{Database Visualization}~\cite{wu2025formalism,Wu2025DatabaseTI}?
This potentially could expand the scope of visualization systems much in
the same way as the introduction of the relational model~\cite{Codd1970ARM}.

\stitle{What About AI?}
The promise of AI is that computing can be personalized to the specific task instance and the user as they are, rather than requiring users to adjust their processes and expectations to what software has implemented. This directly addresses the design dependence discussed above: an interface need not be limited to the designs that its developers anticipated.

However, AI also changes which abstractions are needed. Popular LLM agents such as Claude Code and Codex can already generate visualization designs by writing SVG directly, bypassing visualization libraries altogether. This makes it possible to generate and refine many designs without encoding each through an existing visualization grammar. Similar questions will eventually arise for the underlying data systems. Recent works like GenDB-V~\cite{lao2026gendbv} and Bespoke-OLAP~\cite{Wehrstein2026BespokeOS} synthesize database engines specialized to hardware and workloads, increasingly with machine-checked correctness proofs.  Work on verified code generation in Lean~\cite{sosso2026isomorphism} in growing and applied across computing systems.  Thus, generating correct implementations alone may not be a durable research agenda.

One perspective is that the data processing underlying many basic visualizations is relatively simple---selection, projection, and aggregation---and expressed by variants of OLAP cubes such as imMens~\cite{LiuJH13} and Mosaic~\cite{HeerM24}. While this does not imply that visualization design is ``simple'', but friction due to implementation alone is increasingly tractable through direct synthesis.

\begin{takeawaybox}[float=t]{sec:eugene}
    \begin{takeawayitemize}
		\item
		Co-design of data systems and interfaces is critical: performance, system abstractions, and interaction design   determine user experience, and visualization cannot be optimized independently of data management.

		\item
		Traditional workload-centric metrics (e.g., query latency) are insufficient; interface-aware performance models that account for perceptual constraints, interaction structure, and partial results are necessary to support fluid, interactive analytics.

		\item
        AI agents can increasingly generate visualization interfaces and specialized implementations, but this does not remove the need for system abstractions. Systems must still meet latency and resource constraints, coordinate shared state, enforce data-use policies, and preserve users' expectations across interactions.

	\end{takeawayitemize}
\end{takeawaybox}

What is a durable research direction for interface-system co-design? Agents still rely on tools and execution harnesses that encapsulate recurring logic and provide reliable access to their environment. One direction is to identify requirements that depend on guarantees or runtime capabilities that an agent cannot, or has difficulty, inferring from its observations. For instance, a generated visualization may not meet the latency-resource requirements addressed by Physical Visualization Design~\cite{chen2025physical}, though AI can help navigate the search space.
 Concurrent writes and multiple users require coordination over shared state. Policy and regulatory requirements over data use requires visibility into the data flows hidden behind a SQL tool call, and systems like Data Flow Control~\cite{summers2026dataflow} are needed to enforce them. Finally, evaluating whether an analysis task is even complete may require executing destructive instructions, and new capabilities like database and environment branching are necessary~\cite{xu2026probe,Ang2026BranchBenchAD}.

Finally, users expect properties they may never articulate: consistent interaction behavior, preservation of prior work, or adherence to domain conventions. Personalization requires identifying and maintaining these expectations across tasks and interactions. Clean abstractions that encapsulate complexity therefore likely remain valuable. However, these needs may not directly align with traditional visualization and data management problems. The research challenge is to identify which requirements should be specified, enforced, and exposed by the surrounding system as agents absorb more of the work in reasoning and implementation.

\stitle{\connicon\ Connections to Other Sections.}
Interface-aware optimization can help meet the responsiveness requirements of interactive subset analysis (Sec.~\ref{sec:dominik}) and real-time multimodal guidance (Sec.~\ref{sec:arnab}): exposing view dependencies helps prioritize updates that matter to users.
It also complements progressive data preparation (Sec.~\ref{sec:jean_daniel}), which can provide useful feedback before a collection is fully processed.
Agent-generated interfaces still need these system guarantees, while LLM-assisted exploration raises complementary requirements for provenance-aware verification and appropriate reliance (Sec.~\ref{sec:senjuti}).


 \section{Queries at the Speed of Human Thought}
\label{sec:dominik}

Machine learning (ML) development has shifted from being primarily model-centric to primarily data-centric~\cite{abs-2303-10158}. In the early days of ML, engineers often trained models on hundreds or thousands of data points and meticulously tweaked the modeling function to avoid overfitting. With the popularization of Deep Learning, however, data became increasingly important, and many models were derived from pre-trained models. Today, foundation models have similar transformer architectures, and big quality differences stem from the data on which they are trained or fine-tuned. The data management community is well equipped to tackle many of the quality and quantity challenges \cite{WhangRSL23}. However, the full AIML lifecycle of requirement elicitation \cite{RobertsonWMKH23}, data preparation, monitoring, tuning, and evaluation requires oversight by people who have to slice and dice millions or billions of data records. Since models synthesize patterns, a single record rarely defines the behavior of a model, and therefore ML engineers need to understand patterns and trends in relevant subsets \cite{CabreraFBHTHP23}. Relevant subsets are hard to predict and can rarely be precomputed. ML engineers need to grasp often subtle patterns in large datasets full of information. As Herbert A. Simon puts it, ``What information consumes is rather obvious: it consumes the attention of its recipients. Hence a wealth of information creates a poverty of attention, and a need to allocate that attention efficiently''~\cite{Simon}.
\textit{This need to efficiently allocate attention is where data visualization becomes crucial}.

Data visualization leverages our powerful perceptual systems to let us see patterns and trends in data. It is a critical part of any analysis and a precursor to any rigorous statistical analysis~\cite{tukey_data_1966}. Often, simply looking at the data from various angles uncovers new opportunities for improvement. Data visualization in the AIML lifecycle let us overview and design datasets for training, fine-tuning, and evaluation. It enables the serendipitous discovery of patterns and data issues. At the core of the necessary interactive analysis are interfaces that are fast and fast to use.
Well-designed mixed-initiative systems~\cite{allen99} for analysis make people faster~\cite{WongsuphasawatM16,EppersonGMP24}.
Such systems should balance system-initiated suggestions
with users' ability to steer, modify, or reject
them~\cite{v4_horvitz1999}.
We should continue to invest in good tooling that eliminates barriers to effective analysis.
Yet, many data management systems are not designed for interactive (order-of-millisecond) latencies but rather for seconds or minutes.
\textit{These delays in interactive systems lead to fewer observations made~\cite{LiuH14} and could steer analysts towards convenient and fast data along with all the resulting biases}.
Fast databases,   approximation~\cite{HellersteinHW97,MoritzFD017},
prefetching~\cite{BattleCS16,RahmanAKBKPR17}, and
indexing~\cite{MoritzHH19,MaroulisS22f} can help developers build fast interactive systems. However, they add complexity for tool builders that may prevent quick adaptation of interfaces to new needs. Modern data architectures abstract from low-level optimizations like Mosaic~\cite{HeerM24} and are being rapidly adopted. Yet, many opportunities remain for deeper integration with databases, which are often optimized for order-of-second responses rather than real-time order-of-millisecond responses or may not support many of the encoding transformations used in effective data visualizations like cartographic projections.
A lot of exciting work remains to explore new way to \textit{effectively present billion-record datasets at the speed of human thought to facilitate effective analysis and communication}.

\stitle{\connicon\ Connections to Other Sections.}
The need for responsive subset comparisons connects to interface-aware optimization (Sec.~\ref{sec:eugene}) and progressive or approximate processing (Sec.~\ref{sec:jean_daniel}), which can reduce delays before useful results become available.
The mixed-initiative perspective connects to
guidance-driven multimodal exploration in
Section~\ref{sec:arnab}, where users should be
able to inspect, correct, or reject system suggestions.

\begin{takeawaybox}[float=t]{sec:dominik}
	\begin{takeawayitemize}
		\item
		Nowadays, machine learning development depends heavily on data rather than model tweaks.
		In the AIML lifecycle, data visualization is an effective means for exploring and designing datasets used in training, fine-tuning, and evaluation.

        \item
        Data management systems are often optimized for
        order-of-second responses, whereas interactive analysis
        requires response times aligned with human cognition
        (order-of-millisecond). Delays can reduce observations
        and influence exploratory behavior, potentially
        affecting analytical reasoning.

		\item
		Data management techniques such as approximation, indexing, and prefetching can offer interactiveness   but increase complexity for developers.
		New abstractions and architectures are needed to simplify interfaces adaptation to new requirements.
	\end{takeawayitemize}
\end{takeawaybox}

\section{Scalable Visual Analytics}
\label{sec:jean_daniel}

Visualization is a recognized component of data science, but unlike its sibling domains, such as databases, simulation, and machine learning, its research articles are not as concerned with scalability.
Only 2\% of the articles published at the IEEE VIS conference use a keyword related to scalability, compared to the other domains where scalability and management of large amounts of data concern most of the published articles. Yet, in recent years, several articles have tried to improve the situation, either by describing novel ``scalable'' visualization systems~\cite{HeerM24}, better defining what ``scalability'' means in visualization~\cite{richer2024scalability}, or introducing novel programming paradigms for scaling visualization, such as progressive visual analytics~\cite{hoque2024, stolper2014pva, PDABook} and approximate query processing (AQP)~\cite{KimBPIMR15,ParkCM16,JugelJHM14b}. However, \textit{the big picture of scalability in visualization is not clear yet}. Is there a single, consensual measure to use in determining whether we have achieved scalability in visualization, or several?
What are the next challenges we need to address to fill the scalability gap between visualization and the other data science domains, and is it realistic to believe we will eventually fill that gap?
We identify two key issues to address for achieving scalability in visualization, in addition to handling large amounts of data.

 \stitle{Prepared vs.\ Cold-Started Visualization.}
Visualization systems designed to explore data are interactive and should provide a controlled latency to be usable~\cite{LiuH14}.
But what is an acceptable latency to \emph{start} data analysis?
Most visualization systems advertised for their scalability \textit{do not consider data preparation time a major issue}, assuming that the exploration time will far exceed the preparation time or can be done by someone other than the data analyst.
However, this assumption may no longer be true. With the growing number of public datasets available, finding out which dataset or group of datasets could help answer a question becomes part of the data exploration process~\cite{Auctus}, and the time from deciding to explore a dataset to visualization cannot be considered as \emph{preparation} but should rather be included in the exploration process.
The prepared systems are well-suited to recurring explorations with well-defined queries that can be optimized using indexes. Cold-starting systems are suited to one-time or more open-ended explorations~\cite{UlmerAFKM24, bmpv2021, Alagiannis2012}, such as journalist investigations or discovering datasets and topics within them. The prepared systems correspond to traditional use cases, whereas the cold-starting systems have become more important with the recent profusion of open datasets on the web.
Scalability should now consider \textit{the new use cases made possible by integrating and joining unexplored datasets on-the-fly. Scalable systems should mention their assumptions regarding preparation time when reporting on their scalability.}

Cold-starting visualization comes with several new challenges that are well-suited to the progressive approach but will require more research and technical development to become mainstream since the necessary algorithms and data management systems are not yet ready off the shelf.

\stitle{Scalability of Visualization Techniques.}
Most visualization techniques are not designed to scale to large amounts of data.
When dealing with large-scale data, uncontrollable issues arise, such as large numbers of categories, outliers, and pathological distributions with uneven value densities and possible long tails.
Some of these issues have been documented by Wickham~\cite{wickham2013bin}, but none of the popular visualization libraries provide mechanisms to overcome them. Over the last fifteen years, the visualization community has been relying on the Grammar of Graphics (GoG)~\cite{GrammarOfGraphics} to express visualizations; however, the GoG specifications are not designed to be resilient to big data issues. It would require the visualization programmer to foresee potential data issues and design the right GoG specification before examining the data, which is unrealistic.
When dealing with large amounts of data, visualizations should be designed to be resilient to data issues: applying appropriate aggregations when necessary, filtering out outliers without hiding them completely, and showing aggregated data in dense areas and non-aggregated data in sparse areas.
Building GoG specifications for such operations might be possible, but certainly much more complex than the concise specifications the GoG has been designed for.

Scalable visualization systems should rely on data analysis and hybrid visualizations---such as overlaying outliers onto a density map for scatterplots~\cite{mayorga2013splatterplots}---to adapt to data issues.
Wickham's Bin-summa\-rise-smooth framework describes operators such as ``peel'' to automatically remove low-density extrema from the visualizations. The GoG allows data transformations to better adapt channels to data, such as using a logarithmic transformation for the color scale or for positional axes when the data distribution is not normal. However, these transformations are explicit in the GoG at the specification level.
Scalable visualizations should help analysts in inferring reasonable default values. Scalable visualizations should provide more automation than the GoG allows, and this automation should rely on data analysis and hybrid visualization techniques.

Visualization scalability will not come solely from speeding up algorithms and data structures, or predicting user actions; many paradigms must change.
\textit{Optimizing the traditional visualization pipeline will not scale to the three--to--five orders of magnitude required to keep up with the other data science domains}; progressive and AQP programming should take over. For specifying visualizations, the GoG is too simple to adapt to the unavoidable issues found in large-scale datasets.
\textit{More complex visualization specifications are needed, based on data analysis and adaptation rules}.

\stitle{\connicon\ Connections to Other Sections.}
Progressive and approximate processing can help meet the responsiveness requirements of interactive subset analysis (Sec.~\ref{sec:dominik}), especially when full data preparation would delay the first useful result.
However, when processing is incomplete, the extracted attributes used in exploratory query reformulation (Sec.~\ref{sec:senjuti}) may also be incomplete; systems must distinguish unprocessed records from records that lack a property.

\begin{takeawaybox}[float=t]{sec:jean_daniel}
	\begin{takeawayitemize}
		\item
		Visualization research needs to place greater emphasis on scalability, starting with a clear definition of what scalability entails.

		\item
		Cold-start visualization requires further development to enable real-time exploration of new or large datasets on the fly.

		\item
		Existing visualization techniques must become adaptive and data-aware, automatically handling outliers, skewed distributions, dense regions, and large category counts through aggregation and hybrid representations.

		\item
		New paradigms are needed for scalable visual analytics, including progressive computation and approximate query processing.
	\end{takeawayitemize}
\end{takeawaybox}


\section{Querying Multimodal Data}
\label{sec:arnab}

Interacting with data increasingly goes beyond tables and text and spans multimodal corpora such as images, video, audio, and visually rich documents.
This shift exposes a fundamental challenge in query formulation: \textit{users cannot craft meaningful queries until they understand what the data contains, yet they cannot understand the data without querying it.} For structured data, this chicken-and-egg problem is manageable with a quick exploratory data analysis~(EDA) step or skimming~\cite{singh2012skimmer} through some fragments of the data.

For multimodal data such as video, this bootstrapping is prohibitively expensive. Skimming a video archive is very different from skimming a table -- watching even 1\% of a 10,000-hour video corpus would take several days, and file-level metadata (timestamps, codec, resolution) offer little semantic guidance. Humans become their own bottleneck when consuming such data, exacerbating the query formulation problem.

Here, recent multimodal LLMs and vision-language models (VLMs) enable a crucial unlock: they can ``peek inside'' multimodal content and produce high-quality, zero-shot semantic descriptions.  Rather than expecting users to arrive with well-formed queries, we can now help users discover what they can ask, guiding~\cite{nandi2011guided} them through candidate query concepts using familiar interaction patterns such as autocompletion~\cite{yoo2025emojis, yoo2024guided}. Guidance can be built into complex querying as well, e.g., ZELDA~\cite{romero2023zelda} uses VLMs to extract event semantics from surveillance footage, enabling queries like ``show me unusual pedestrian behavior'', and SketchQL~\cite{wu2024sketchql} allows users to sketch what they're looking for, and VLMs match it against frames. Query synthesis approaches~\cite{10.14778/3611479.3611482} infer compositional \mbox{video-event} queries from limited user interactions, reducing the need for expertise in declarative query languages. These systems shift the interaction model from \textit{query--then--explore} to \textit{explore--with--guidance}, where the system continuously suggests what is discoverable as users navigate. However, making VLM-driven guidance practical for interactive visual analytics exposes challenges across database systems, HCI, and visualization.

\stitle{Trust and Feedback.} These unlocks enabled by multimodal LLMs and VLMs come hand-in-hand with hallucination and non-determinism challenges~\cite{fu2025mitigating}. Guidance provided to the user is inherently brittle and may erode trust when below a quality threshold. For instance, a VLM might confidently tag a video frame as containing ``bicycles'' when the image actually shows motorcycles, potentially misleading the user's data exploration. Unlike text where users can quickly read an answer to verify correctness, validating multimodal suggestions is expensive; users cannot watch every suggested clip. Systems must provide fast verification mechanisms to ameliorate such issues, e.g., thumbnails or ``highlights reels''~\cite{winecki2024v2v} rather than requiring full video playback, or risk users abandoning guidance due to high verification costs that rival manual exploration.
Beyond verification, provenance and feedback mechanisms
should allow users to inspect, correct, or reject system
suggestions~\cite{v4_horvitz1999}, supporting appropriate
reliance rather than uncritical trust~\cite{LeeSee2004}.
Alternatively, the system itself can monitor its suggestions, and self-reflect on possible mistakes as a self-improvement step. These concepts are aligned with the interaction, lineage, and feedback components discussed in prior sections.

\stitle{Next-Generation Interfaces Amplify the Need for Guidance.} As interfaces move toward augmented
and virtual reality (AR/VR) and ``in-the-world'' querying~\cite{burley2019arquery,khan2021dreamstore}, guidance becomes even more critical. Users wearing AR headsets in warehouses or medical settings cannot type complex queries; they need systems that understand gestural or spoken intent and immediately suggest what's discoverable. Consider a warehouse worker asking ``show me damaged pallets'' while looking at inventory. The system must process live video, recognize relevant objects, and suggest subclauses (``damaged on left side?'' ``within last week?'') in real-time, guiding the user to the right questions to ask. Camera-first interaction paradigms~\cite{wolf2023camera} and gestural query specification~\cite{nandi2013gestural} require query abstractions that accept multimodal inputs (sketches, gestures, exemplar images, natural language) and provide real-time feedback~(e.g., via AQP~\cite{jo2024thalamusdb}) on what's queryable, or risk disorientation within such highly interactive interfaces. In settings such as augmented reality, guidance also comes with hard latency and performance requirements since the user is interacting within the real-world and mistakes can be safety-critical.

\stitle{Representing Guidance.}  Making guidance effective for multimodal exploration requires addressing how to shape VLM outputs for human consumption under tight interaction budgets. The challenge is not just inference speed, but usability: textual descriptions of visual content are verbose and cognitively difficult to skim quickly. A VLM might return ``a person wearing a red jacket riding a bicycle on a paved path near trees during daytime'' for a single frame. While this information is accurate, it is difficult for a human to process when scanning hundreds of suggestions. Effective guidance needs \emph{succinct, readable visual cues}, possibly in the form of icons, thumbnails, or even emoji-like glyphs~\cite{yoo2025emojis} that let users quickly assess what a suggestion means and whether it matches their intent. This also raises fundamental questions about \textit{guidance granularity}: Should systems propose coarse concepts (e.g., \textit{``sports,''}) that characterize the overall content of the data, but risk being too repetitive across queries, or fine-grained evidence (\textit{a specific thumbnail showing a bicycle at timestamp 1:23}) that points to concrete queries? What level of specificity best helps users formulate their first query without overwhelming them with detail?

Addressing these challenges requires cross-disciplinary collaboration, bridging concepts in query optimization, representation design, mixed-initiative interaction, and progressive visualization to build systems that   guide users through the
 query formulation barrier in multimodal analytics.

\begin{takeawaybox}[float=t] {sec:arnab}
	\begin{takeawayitemize}
		\item
        Multimodal data such as video collections present a query formulation barrier: users cannot query what they don't know exists, yet they cannot discover what's queryable without prohibitively expensive manual inspection.

		\item
        VLMs unlock this dependency, enabling guidance-driven exploration by proposing zero-shot query suggestions and autocompletion-style querying based on semantic understanding of content. This shifts interaction from \textit{query-then-explore} to \textit{explore--with--guidance}.

		\item
        Making guidance practical requires efficient
        verification and user control over system suggestions.
        For next-generation interfaces (AR/VR, gesture,
        camera-first), guidance must also be timely and
        represented through succinct visual cues to support
        rapid exploration.
    \end{takeawayitemize}
\end{takeawaybox}

 \stitle{\connicon\ Connections to Other Sections.}
Guidance for unfamiliar multimodal collections helps users
formulate initial queries, complementing iterative query
reformulation and result selection (Sec.~\ref{sec:senjuti}).
Source-linked visual cues support verification, while
mixed-initiative guidance requires timely suggestions
and user control (Sec.~\ref{sec:dominik}), along with
interface--system co-design (Sec.~\ref{sec:eugene}).


\section{\mbox{Interactive Data Exploration}}
\label{sec:senjuti}

Interactive data exploration plays a pivotal role in discovery-driven applications, where the process of data discovery is ad hoc and highly interactive. In these applications, users write multiple queries with varying criteria, aiming to strike a balance between gathering all relevant information and avoiding an overwhelming volume of data.  The data management community has closely investigated the need to help users reformulate their queries during data exploration so that the retrieved sets are \textit{not too large} (many answer problems)~\cite{basu2008minimum}, \textit{not too small} (empty answer problem)~\cite{mottin2013probabilistic}; or are \textit{not off-target and diverse} (ensuring relevance and diversity)~\cite{islam2023generic,nikookar2023diversifying}.

When data is structured, existing work has leveraged metadata or attributes to understand the user's intent and seek their preferences accordingly. Algorithms were designed to navigate users to serendipitous, relevant, diverse representatives of data or to help users find the entity of interest while minimizing effort, e.g., minimize the number of clicks users provided.  These algorithms aided users to reformulate their   queries so that users could interactively explore related data before landing on a small sample of their final interests.

AI has led to an unprecedented explosion in the volume and complexity of unstructured data. With the rise of unstructured data, ranging from text and images to video and audio, the need for interactive ad hoc data exploration has never been greater. Users frequently engage in query reformulation and seek relevant, diverse~\cite{vieira2011query}, and sometimes serendipitous information. However, the lack of structured metadata, which traditionally guided exploration, presents a significant challenge. As AI continues to reshape the landscape of data exploration, the following question therefore arises: ``\textit{Can we still rely on traditional data exploration tools to navigate and uncover insights from the increasingly complex data}?''.

The classical data exploration techniques can still make a headway in this new landscape, provided there is a seamless integration between the structured and unstructured data worlds. Identifying structures from unstructured data requires a combination of advanced techniques in NLP, machine learning, image and speech recognition, and data integration~\cite{shah2020speakql}. LLMs are particularly useful, too, as they excel at understanding context, extracting meaning, and processing data of different modalities including audio and video. \textit{Although LLMs have demonstrated impressive capabilities in processing unstructured data, several key research challenges still prevail}.  Non-determinism of LLMs is one such key challenge refers to the inherent unpredictability or variability in the output generated by these models, even when the same input is provided multiple times.  This introduces a level of unpredictability, requiring the data exploration algorithms to be reconsidered through a \textit{probabilistic lens}~\cite{suciu2022probabilistic}.
Other challenges include the high cost of invoking these models~\cite{nia2025personalized}, their potential biases~\cite{islam2023equitable}, limited contextual understanding, inability to adapt to highly specialized domains, and significant computational resource demands.

 Beyond these technical challenges, LLM-driven data exploration raises fundamental questions about human sensemaking, cognitive trust, and appropriate reliance. Data exploration is inherently an iterative sensemaking process in which users formulate hypotheses, interpret emerging patterns, and refine their understanding~\cite{v4_russell1993,v4_pirolli2005}. While LLMs can facilitate this process, their generated insights may be difficult to verify or trace to the underlying evidence. This calls for provenance-aware and verifiable exploration methods that connect generated insights to their supporting data, communicate uncertainty, and enable users to assess their validity~\cite{v4_cassenti2023}. Equally important is supporting users in developing accurate mental models of LLM capabilities and limitations, particularly given the cognitive challenges of interacting with these models~\cite{v4_subramonyam2024}. Explainable and human-centered interfaces can further support users in interpreting AI-generated insights~\cite{v4_bernardo2023}. Ultimately, these capabilities are essential for calibrating cognitive trust and promoting appropriate reliance, ensuring that users neither over-rely on plausible but incorrect outputs nor underutilize reliable insights~\cite{LeeSee2004}.

To overcome these challenges, it is crucial to complement LLM-driven processes with expert human oversight~\cite{AmershiWVFNCSIB19}, careful data curation~\cite{guideline2}, and transparency in model training and usage~\cite{guideline3}. Overall, this is an exciting time for data researchers to collaborate, bringing together expertise across fields to create more nuanced and adaptable solutions for the evolving challenges of data exploration.

\vspace{3pt}
\noindent\textbf{\connicon} \textbf{Connections to Other Sections.}
Multimodal guidance helps users identify initial query concepts (Sec.~\ref{sec:arnab}), whereas the exploration processes discussed here focus on reformulating queries and assessing results as analysis proceeds.
Errors or gaps in model-extracted attributes can affect selected records and downstream narratives (Sec.~\ref{sec:yifan}); provenance-aware verification is therefore important for appropriate reliance. Agent-generated interfaces (Sec.~\ref{sec:eugene}) introduce separate requirements for consistent interaction behavior.

\begin{takeawaybox}[float=t]{sec:senjuti}
\begin{takeawayitemize}
 \item
The growth of unstructured and multimodal data challenges traditional metadata-driven exploration, increasing the need for interactive, ad hoc analysis and continuous query reformulation.

\item
LLM-assisted exploration introduces distinct failure modes, including hallucinations, non-determinism, bias, and prompt sensitivity. Addressing them requires verification and mitigation strategies, while different model integration approaches introduce trade-offs in reliability, verification, and cost.

\item
Trustworthy AI-assisted exploration requires supporting human sensemaking and appropriate reliance through provenance, verifiable insights, uncertainty communication, explainable interfaces, accurate mental models of LLM capabilities, and human oversight.
\end{takeawayitemize}
\end{takeawaybox}

\section{Aesthetic and User Guidance in Generative Data Visualization}
\label{sec:yifan}

Information visualization enables the graphical representation of data and information that helps users find patterns, trends, and relationships within the data, and to communicate complex data sets clearly and effectively.
In the field of network visualization, traditionally~\cite{eades84,KamadaK89,GansnerKN04}, visualization of undirected graphs relies on heuristics, especially those based on modeling graphs as physical systems, with the belief that minimizing the energy of such systems leads to aesthetically pleasing layouts that help illustrate the underlying unstructured data.
However, it has not been proven that human aesthetic preference can be well modeled by such physical systems. With the advent of DL, there is a growing interest in taking advantage of the power of neural networks to help visualize network data, e.g.,~\cite{WangJWCMQ20,KwonM20}.
In particular, it is now possible to use DL to expand the horizon of traditional graph visualization in a number of directions.

It was demonstrated that a Graph Neural Network (GNN) model can be used to optimize arbitrary differentiable objective functions.
Once trained, such a model can be applied to arbitrary graphs never seen in the training data and create visualizations that optimize the objective function even better than traditional benchmark algorithms~\cite{WangYHS21,TiezziCG24,GiovannangeliLAGB24}.
Recently, it was shown that by using the Generative Adversarial Network (GAN), we can even train neural networks to optimize non-smooth objective functions~\cite{WangYHS24}.
In fact, this approach does not require access to the objective function at all and only requires a comparative function that can choose between two visualizations the ``better'' one (based on a hidden objective function unknown to GAN).

This opens avenues for future investigation: \textit{instead of optimizing arbitrary objectives such as edge crossing or stress, we should be able to model the human sense of aesthetics directly}.
For example, if reducing edge crossing is most important for humans, Tiezzi et al.~\cite{TiezziCG24} demonstrated that a fully connected network can classify whether two edges cross; on the other hand, Wang et al.~\cite{WangYHS24} have shown that the discriminator in a GAN setup can be trained to choose drawings with a lower number of edge crossings, learning from pairs of bad and good drawing examples.
The next step is to train a model on a collection of human-labeled example visualizations to learn human visual preference.
We believe such a system is possible, but the challenges include scaling GNN, GAN, or other DL-based models to very large unstructured data, and scaling human label collection.
Some progress in scalability has already been made~\cite{YanZY22a,GrotschlaMVW24,HuangZXLZ21}, including recent work which demonstrated that VLMs, and to a lesser extent LLMs, can be trained to align with human easthetic preferences, thus acting as viable human proxies for label collection~\cite{Li2026Beauty}.
More work is needed 
make network visualizations adapt to personalized human preferences, incorporate human-specified constraints, and run   faster than traditional network layout algorithms.
However, aesthetic preference alone does not establish
analytical usefulness. Evaluation should also consider
how well learned layouts support specific visual analysis
tasks.

Another future direction involves \textit{not just producing the visualization itself but also crafting a narrative, or even combining it with animations/videos} that elucidate the patterns depicted within the visualization, thus directing the user's attention to the most crucial aspects of generated visualization.

Accomplishing this task will \textit{necessitate leveraging multimodal LLMs and VLMs trained extensively on a diverse corpus consisting of source data (e.g., CSV files, unstructured data), their visualizations, and their corresponding captions} extracted from past visualization and other scientific literature.
If we can solve the above challenges, we can then apply such a model to any data to be analyzed, and automatically generate sample visualizations, animations, and narratives that are not only aesthetically pleasing to look at, but also informative, with a well-told story about the visualization.
Generated narratives should also be evaluated for
faithfulness to the underlying data, including omissions
that may result from upstream data selection.

  \begin{takeawaybox}[float=t]{sec:yifan}
 	\begin{takeawayitemize}
 		\item
 		Deep learning techniques can significantly advance graph visualization by optimizing layouts and learning human aesthetic preferences beyond traditional metrics such as edge crossings.

 		\item
 		Training aesthetic deep learning models introduces major scalability challenges when large-scale training datasets are involved. Additionally, collecting sufficient human preference data remains a significant challenge.

 		\item
 		Visualization systems should generate not only layouts but also narratives and animations that highlight crucial data patterns and guide user attention. Achieving this requires multimodal models trained on datasets that include both visualizations and explanatory context, such as captions.
 	\end{takeawayitemize}
 \end{takeawaybox}

\stitle{\connicon\ Connections to Other Sections.}
Learning-based layouts share the scalability concerns of large-data visualization (Sec.~\ref{sec:jean_daniel}), although training and inference introduce different costs from progressive processing. Agent-generated interfaces (Sec.~\ref{sec:eugene}) introduce separate questions about system guarantees.
Generated narratives also depend on the completeness and reliability of records selected during exploration (Sec.~\ref{sec:senjuti}): without provenance-aware verification, an aesthetically convincing account may still omit relevant information if upstream selection is incomplete.


\section{Discussion and Insights}
\label{sec:disc}

In this section, we discuss how the concrete problems and research opportunities identified throughout the paper relate to the main challenges outlined in Table~\ref{tab:shared-challenges}, and further analyze how these challenges interact.

\subsection*{C1. Responsiveness}
\vspace{4pt}

\textbf{C1. Challenge Aspects.}
Responsiveness extends beyond the latency of an individual query. A single interaction may trigger several linked updates, requiring the interface to determine which views must update together, which can be deferred, and which responses matter most (Sec.~\ref{sec:eugene}). Delayed feedback also disrupts analysis as users move between individual records and dynamically defined subsets of large training, fine-tuning, and evaluation datasets, while specialized optimization mechanisms may limit adaptability (Sec.~\ref{sec:dominik}). The time required to prepare unfamiliar datasets for analysis is another part of the problem (Sec.~\ref{sec:jean_daniel}). In spoken, gestural, camera-first, and AR/VR interactions, feedback must arrive while users act within their environment (Sec.~\ref{sec:arnab}).

\stitle{C1. Opportunities.}
Exposing interface structure, update dependencies, and response requirements to the optimizer offers an alternative to treating the interface as a conventional query workload (Sec.~\ref{sec:eugene}). Approximation, prefetching, and indexing can better support exploratory tools (Sec.~\ref{sec:dominik}), while progressive computation and approximate query processing (AQP) can expose useful results while preparation continues (Sec.~\ref{sec:jean_daniel}). For multimodal interactions, processing must also provide timely feedback about which data and operations are available for interaction (Sec.~\ref{sec:arnab}).

\stitle{C1. Cross-Challenge Interactions \& Synthesis.}
Responsiveness is not only a backend performance concern; it depends on how computation, data preparation, and interaction are coordinated. For instance, subset comparisons (Sec.~\ref{sec:dominik}) may trigger updates across multiple linked views, making interface dependencies essential for deciding which results must arrive together (Sec.~\ref{sec:eugene}). Progressive preparation can shorten the time to first analysis (Sec.~\ref{sec:jean_daniel}), but partial or sampled results may overlook rare model failures of interest (Sec.~\ref{sec:dominik}). Early feedback is therefore valuable only when users can understand which data and computations it represents. Specialization can also limit adaptability: precomputation, indexing, and workload-specific execution can accelerate anticipated interactions while constraining support for unanticipated analytical paths. To assess responsiveness, we need to consider when results become available, the evidence they provide, their completeness, and their suitability for the task.

\subsection*{C2. Scalability}
\vspace{4pt}

\textbf{C2. Challenge Aspects.}
Scalability involves different types of cost. Data preparation is part of the problem, as are the difficulties that dense regions, outliers, and high-cardinality attributes pose for fixed visualization specifications (Sec.~\ref{sec:jean_daniel}). Selecting and composing specialized optimizations is also challenging, particularly when relational dependencies are obscured by single-table assumptions (Sec.~\ref{sec:eugene}). Interactive comparisons must be preserved over very large training and evaluation datasets, even when the relevant subsets are defined dynamically during analysis (Sec.~\ref{sec:dominik}). For video and multimodal collections, even manual inspection of a small fraction can be costly (Sec.~\ref{sec:arnab}), while repeatedly invoking models to extract or predict attributes adds computational cost to exploration (Sec.~\ref{sec:senjuti}). Learned graph layouts introduce further requirements for sufficient training data and computationally practical models as graph sizes and learned-layout workloads grow (Sec.~\ref{sec:yifan}).

\stitle{C2. Opportunities.}
These costs can be addressed at different stages. Progressive processing and AQP can be combined with explicit accounting of preparation cost (Sec.~\ref{sec:jean_daniel}). Interface-aware optimization and richer relational execution (Sec.~\ref{sec:eugene}) are complemented by abstractions that make approximation, prefetching, and indexing accessible within exploration tasks (Sec.~\ref{sec:dominik}). For multimodal processing, approximation can avoid unnecessary computation (Sec.~\ref{sec:arnab}), while extracted attributes can be reused with their sources and extraction versions tracked (Sec.~\ref{sec:senjuti}). Preference-aware layout learning offers another direction for scaling with growing datasets and model workloads (Sec.~\ref{sec:yifan}).

\stitle{C2. Cross-Challenge Interactions \& Synthesis.}
Scalability often involves shifting computational costs between stages rather than eliminating them. Indexing, precomputation, attribute extraction, and model training can accelerate subsequent interactions, but introduce preparation, storage, and refresh costs. Progressive processing illustrates these dependencies. When retrieval continues during exploration (Sec.~\ref{sec:jean_daniel}), attribute-based methods may operate on metadata covering only part of the collection (Sec.~\ref{sec:senjuti}). A missing label may therefore indicate that a record has not yet been processed, rather than that the corresponding property is absent. Treating these cases equivalently can incorrectly exclude relevant records. Learned visualization raises a related concern: scaling these methods requires not only sufficient computational resources, but also adequate training data and evidence that learned behavior generalizes beyond the training workload (Sec.~\ref{sec:yifan}). Scalability claims should consequently specify which computations are performed in advance and distinguish preparation, refresh, inference, and interactive execution costs.

\subsection*{C3. Guidance \& User Control}

\vspace{4pt}
\textbf{C3. Challenge Aspects.}
Guidance and user control matter at different stages of exploration. At an early stage, users may not yet know what a multimodal collection contains, and model-generated concepts may overlook useful ways to explore it (Sec.~\ref{sec:arnab}). An AI-generated interface may propose joins, aggregations, and update behavior, requiring users to inspect and revise the intended operation rather than accept opaque generated views (Sec.~\ref{sec:eugene}). As analysis proceeds, relevant comparisons and subsets emerge that cannot all be anticipated by predefined dashboards (Sec.~\ref{sec:dominik}). Queries that return too few, too many, or irrelevant results require iterative refinement, with users balancing relevance, diversity, serendipity, and interaction effort (Sec.~\ref{sec:senjuti}). Control also extends to user-specified layout constraints and narratives or animations designed to direct attention (Sec.~\ref{sec:yifan}).

\stitle{C3. Opportunities.}
Several mechanisms can help keep users involved in exploration. Autocompletion and multimodal query inputs can be combined with user-defined concepts and suggestions linked to their source evidence (Sec.~\ref{sec:arnab}). Generated operations can be made inspectable and revisable, and personalized interfaces should preserve users' expectations across interactions (Sec.~\ref{sec:eugene}); subset definitions can remain editable, with mixed-initiative suggestions about what to investigate (Sec.~\ref{sec:dominik}). Query reformulation and result selection can also be extended to model-extracted attributes, with support for human sensemaking and accurate mental models of LLM capabilities (Sec.~\ref{sec:senjuti}). For visual representations, users can specify layout constraints and guide attention through narratives and animations (Sec.~\ref{sec:yifan}).

\stitle{C3. Cross-Challenge Interactions \& Synthesis.}
Guidance and user control become interdependent when multiple stages of exploration rely on the same derived evidence. For example, if video labels inform both query suggestions (Sec.~\ref{sec:arnab}) and result selection (Sec.~\ref{sec:senjuti}), a motorcycle mislabeled as a bicycle may lead users toward an inappropriate query and cause the corresponding clip to appear among bicycle results. Correcting the source label is insufficient unless the correction propagates to dependent suggestions and result sets. More generally, the form of control users need depends on where they are in the exploration process: they may need to revise a generated operation, introduce a missing concept, edit a subset definition, reject a suggestion, or reformulate a query.

These mechanisms must also balance guidance against interaction
effort. Classical sensemaking models frame this trade-off
in terms of information acquisition and organization
costs~\cite{v4_russell1993}, and the interplay between
information-foraging and sensemaking
loops~\cite{v4_pirolli2005}.
Broad concepts may produce repetitive or uninformative suggestions,
whereas fine-grained evidence can be more costly to inspect.
Effective guidance therefore depends not simply on the quantity
of assistance provided, but on its relevance, timing, traceability,
and the user's ability to evaluate and override it.

\subsection*{C4. Reliability \& Appropriate Reliance}
\vspace{4pt}

\textbf{C4. Challenge Aspects.}
Reliability concerns both the correctness of analytical operations and the limitations of AI-assisted processing. Concealing join dependencies behind a single-table view can introduce aggregation errors; in AI-generated interfaces, successful execution likewise does not guarantee that the generated joins and aggregations reflect the intended analysis (Sec.~\ref{sec:eugene}). Model suggestions may be wrong or unstable and expensive to verify when the evidence is video rather than text (Sec.~\ref{sec:arnab}), while missing, uncertain, variable, or biased extracted attributes can change which records are selected (Sec.~\ref{sec:senjuti}). For learned visualizations, alignment with human aesthetic preferences does not necessarily imply analytical effectiveness, and generated narratives must accurately reflect the values and relationships in the underlying data (Sec.~\ref{sec:yifan}).

\stitle{C4. Opportunities.}
Several directions make errors easier to detect and assess. Query and provenance abstractions can expose generated operations for inspection (Sec.~\ref{sec:eugene}). Source-linked thumbnails or highlight reels can reduce the effort required to verify multimodal suggestions, while feedback and correction can be propagated through the system (Sec.~\ref{sec:arnab}). For model-extracted attributes, recording source and version information, supporting uncertainty-aware selection, and checking for omitted relevant records can clarify what has been established and what remains uncertain (Sec.~\ref{sec:senjuti}). Evaluating learned visualizations similarly requires distinguishing preference alignment, analytical task performance, and narrative faithfulness (Sec.~\ref{sec:yifan}).

These AI-specific failure modes need to be addressed differently. Hallucinated labels or narrative claims should be checked against source evidence or computed results. Hallucination-targeted preference optimization offers one possible model-level intervention in specific multimodal settings~\cite{fu2025mitigating}, but does not provide a general guarantee of correctness. Non-deterministic outputs require recording accepted results and the settings that produced them, with repeated-run comparisons where appropriate. Bias requires examining systematic errors and uneven coverage, including omissions across relevant groups. Prompt sensitivity presents a related but distinct challenge: even deterministic models may produce different results when prompt wording, examples, or their ordering changes~\cite{v4_zhao2021}.

The choice of model integration strategy also affects what can
be verified and where costs arise.
Prompting supplies instructions or examples without task-specific
parameter updates~\cite{v4_brown2020}.
Retrieval-augmented generation (RAG) introduces an explicit
retrieval stage~\cite{v4_lewis2020}, which can make retrieved
evidence available for inspection when it is retained and exposed.
However, its effectiveness depends on the relevance and coverage
of retrieved content, and retrieval may add processing costs.
Fine-tuning, including parameter-efficient approaches such as
low-rank adaptation~\cite{v4_hu2022}, changes model parameters
and requires tracking training data and model versions, but does
not necessarily provide evidence for individual outputs.
These strategies can be combined, but they differ in computational
cost, sources of uncertainty, and support for verification.

\stitle{C4. Cross-Challenge Interactions \& Synthesis.}
Reliability failures can propagate across successive stages of exploration. Progressive or sampled processing may leave rare failures outside the current view (Sec.~\ref{sec:dominik}~\&~\ref{sec:jean_daniel}). Incomplete attribute extraction may cause an unknown value to be interpreted as a negative one, changing which records are displayed (Sec.~\ref{sec:jean_daniel}~\&~\ref{sec:senjuti}). An incorrect multimodal label may then affect both query suggestions and result selection unless a reviewed correction reaches every dependent operation (Sec.~\ref{sec:arnab}~\&~\ref{sec:senjuti}). These errors can also affect downstream visualizations: a narrative may accurately describe the records it receives even though those records already exclude relevant items (Sec.~\ref{sec:senjuti}~\&~\ref{sec:yifan}). Provenance must therefore capture more than data origin. Coverage, extraction status, versions, and dependencies are also needed to determine what has been verified, what remains uncertain, and which downstream results require updating after a correction. Agent-generated workflows may additionally require coordination of shared state, data-use policy enforcement, and safe branching for potentially destructive operations (Sec.~\ref{sec:eugene}).

The objective is appropriate reliance, not simply greater trust in AI. Users need sufficient evidence to recognize important limitations and decide when to verify, reject, or revise an output.
This perspective aligns with research on trust in imperfect
automation~\cite{LeeSee2004}.
Explanation design may influence perceived credibility
and affective responses without necessarily improving the
correctness of AI-generated results~\cite{v4_bernardo2023}.
Users' mental models and expectations, as well as how
uncertainty is presented, may also affect their interpretation
of AI-assisted results~\cite{v4_subramonyam2024,v4_cassenti2023}.
However, explanations and uncertainty indicators do not
guarantee that users will recognize consequential errors.
Evaluation should therefore consider whether users can
identify such errors and make informed decisions,
rather than simply whether system limitations are exposed.

\subsection*{C5. Effective Visual Representation}
\vspace{4pt}

\textbf{C5. Challenge Aspects.}
What constitutes a useful representation varies across the six perspectives. Single-table visualization assumptions can obscure relational structure; although AI agents can generate interfaces beyond fixed visualization grammars, their correctness and interaction behavior still depend on system support (Sec.~\ref{sec:eugene}). Comparisons over large datasets require views that connect model outputs, individual records, and editable subsets (Sec.~\ref{sec:dominik}), while skew, outliers, dense regions, and high-cardinality attributes challenge fixed visualization specifications (Sec.~\ref{sec:jean_daniel}). Verbose descriptions of visual content are difficult to scan when evaluating many suggestions (Sec.~\ref{sec:arnab}). For learned graph layouts, the challenge is to accommodate human preferences and explicit constraints, while narratives and animations need to guide attention without making unsupported claims (Sec.~\ref{sec:yifan}).

\stitle{C5. Opportunities.}
Richer query and visualization abstractions can expose relational structure (Sec.~\ref{sec:eugene}), while responsive views can support editable subsets (Sec.~\ref{sec:dominik}). Aggregation, transformations, and hybrid views, such as density maps that preserve the visibility of important outliers, offer ways to handle difficult data distributions (Sec.~\ref{sec:jean_daniel}). When presenting suggestions, lengthy descriptions can be replaced with compact, evidence-linked cues such as icons, thumbnails, or short clips (Sec.~\ref{sec:arnab}). Preference-aware and constraint-aware layout learning can also be combined with narratives and animations grounded in the underlying data (Sec.~\ref{sec:yifan}).

\stitle{C5. Cross-Challenge Interactions \& Synthesis.}
Visualization fidelity cannot be captured by a single criterion. Aggregation and hybrid views aim to make dense data interpretable without obscuring important outliers (Sec.~\ref{sec:jean_daniel}), whereas learned layouts focus on accommodating human preferences and user-defined constraints (Sec.~\ref{sec:yifan}). In both cases, however, the representation must support the analytical task: visual clarity or aesthetic appeal alone does not establish that users can identify the patterns and relationships they look for.

The quality of the representation also depends on the completeness and reliability of its inputs. If result selection excludes relevant records because extracted attributes are missing or incorrect (Sec.~\ref{sec:senjuti}), a narrative may faithfully describe its input while providing an incomplete account of the collection (Sec.~\ref{sec:yifan}). Likewise, compact cues reduce the effort of scanning suggestions only if they preserve sufficient evidence for users to assess their relevance and verify their content (Sec.~\ref{sec:arnab}). We therefore need to consider both how information is presented and which analytical structures, relevant records, and supporting evidence the representation preserves.

\bibliographystyle{ACM-Reference-Format}
\bibliography{bibliography}

\end{document}